\documentclass[aps,prl,twocolumn,superscriptaddress]{revtex4-1}
\usepackage{amsmath}
\usepackage{amssymb}
\usepackage{amsthm}
\usepackage{mathtools}
\usepackage{braket}
\usepackage{bm}
\usepackage{bbm}
\usepackage{graphicx}
\usepackage{xcolor}
\usepackage{bbold}
\usepackage{blindtext}
\usepackage{graphicx}
\usepackage{graphics}
\usepackage{verbatim}   
\usepackage{amsfonts}
\usepackage{adjustbox}

\newcommand{\numberset}{\mathbb}

\newcommand{\Z}{\numberset{Z}}

\newcommand{\bx}{\mathbf{x}}
\newcommand{\bC}{\mathbf{C}}

\newcommand{\be}{\mathbf{e}}
\newcommand{\by}{\mathbf{y}}

\newcommand{\bQ}{\mathbf{Q}}
\newcommand{\rmi}{\mathrm{i}}

\newcommand{\bna}{\boldsymbol{\nabla}}

\usepackage{siunitx}

\usepackage[french]{babel}
\usepackage[T1]{fontenc}
\usepackage[utf8]{inputenc}

\begin{document}

\title{ Landau theory of crystal plasticity}
\date{\today}

\author{R. Baggio}
\affiliation{CNRS, LSPM, Université Paris 13,   France}
\affiliation{PMMH, ESPCI, Paris, France}
\author{E. Arbib}
\affiliation{Department of Physics, Politecnico di Milano,   Italy}
\author{P. Biscari}
\affiliation{Department of Physics, Politecnico di Milano,   Italy}
\author{S. Conti}
\affiliation{Institut fur Angewandte Mathematik, Universit\"at  Bonn, Germany}
\author{L. Truskinovsky}
\affiliation{PMMH, ESPCI, Paris, France}
\author{G. Zanzotto}
\affiliation{DPG, Università di Padova, Italy}
\author{O.U. Salman}\email{Corresponding author: umut.salman@polytechnique.edu}
\affiliation{CNRS, LSPM, Université Paris 13,   France}


\begin{abstract}
We show that nonlinear continuum elasticity  can be effective in modeling  plastic  flows in crystals  if  it is viewed as  Landau theory with an infinite number of equivalent energy wells whose configuration is dictated by the symmetry group $\mathrm{GL}$ $(3,\Z)$. Quasi-static loading can be then handled by athermal dynamics,  while lattice based discretization can  play the  role of regularization.  As a proof of principle  we study  in this Letter  dislocation nucleation in a homogeneously sheared 2D crystal and   show  that   the global tensorial   invariance  of the elastic energy  foments  the development of  complexity in the  configuration of  collectively nucleating defects. A crucial role in this process is played by the unstable  higher  symmetry crystallographic phases, traditionally thought to be unrelated to plastic flow in  lower  symmetry lattices.

\end{abstract}


\pacs{PACS list, to be added}


\maketitle
Crystal plasticity is the simplest among  yield  phenomena in solids \cite{Bonn2017-er}, and yet it has been compared in complexity to fluid turbulence \cite{Cottrell2002-dh,choi2012dislocation}.  The intrinsic irregularity of   plastic flow   in crystals \cite{Vinogradov2018-gt} is due to  short and long range interaction of  crystal  defects  (dislocations) \cite{Zepeda-Ruiz2017-gu} dragged by the applied loading through a  rugged energy landscape \cite{SALMAN2012219,kubin2013dislocations,ispanovity2014avalanches}.  Fundamental understanding of plastic  flow in crystals is  crucial for  improving   hardening  properties  of  materials \cite{Hughes2018-ge},   extending their  fatigue life \cite{Irastorza-Landa2016-ir},  controlling their  forming at sub-micron scales \cite{Csikor2007-ua} and  building new  materials \cite{McDowell2018-vv}.  

Macroscopic  crystal plasticity  relies  on a phenomenological  continuum description  of plastic deformation in terms of  a finite number of  order parameters representing  amplitudes of pre-designed \emph{mechanisms}.   These mechanisms  are  coupled  elastically   and operate according to   friction type   dynamics
 \cite{Roters2010-cw,Forest2019,Gurtin2009-bl,Han2005-uh,Stein2018-gr}.   The alternative microscopic  approaches, relying  instead  on  molecular dynamics   \cite{Juan1993-cu,bulatov1998connecting,Lyu2019-no,McDowell2018,Moretti2011-xw, Skaugen2018-tb,Tarp2014-ro,Elder2002-qt}, can handle  only  macroscopically  insignificant  time and length scales \cite{McDowell2019-jl}.  An  intermediate discrete dislocation dynamics   approach focuses on  long range interaction  of few dislocations,  while their  short range interaction is still treated  phenomenologically  \cite{Shilkrot2004-ek,Song2019-cc,El-Awady2016-ea}. Collective dynamics of many  dislocations  can be also  described by the  dislocation density field, however, rigorous    coarse-graining in such strongly interacting system  still remains a  major challenge \cite{El-Azab2000-wo,Xia2015-vo,Groma2019-gq,Acharya2006-nm,Chen2013-vd,Sandfeld2011-sc,LeSar2014-wb,Valdenaire2016-yp}. 
 
  A   highly successful computational bridge between microscopic and macroscopic  approaches is provided by the  quasi-continuum  finite element  method  which uses   adaptive meshing  and  employs ab initio approaches   to guide the constitutive response at different mesh scales \cite{Tadmor1996-qi,Tadmor2017-lk,Kochmann2016-bv,Li2002-uj,Li2004-vb}.  Its drawbacks, however,  are spurious effects due to matching of   FEM   representations at different scales and a high computational cost of reconstructing the constitutive response at the smallest  scales \cite{Luskin2013-pe}.
 
\begin{figure}
\includegraphics[scale=0.15 ]{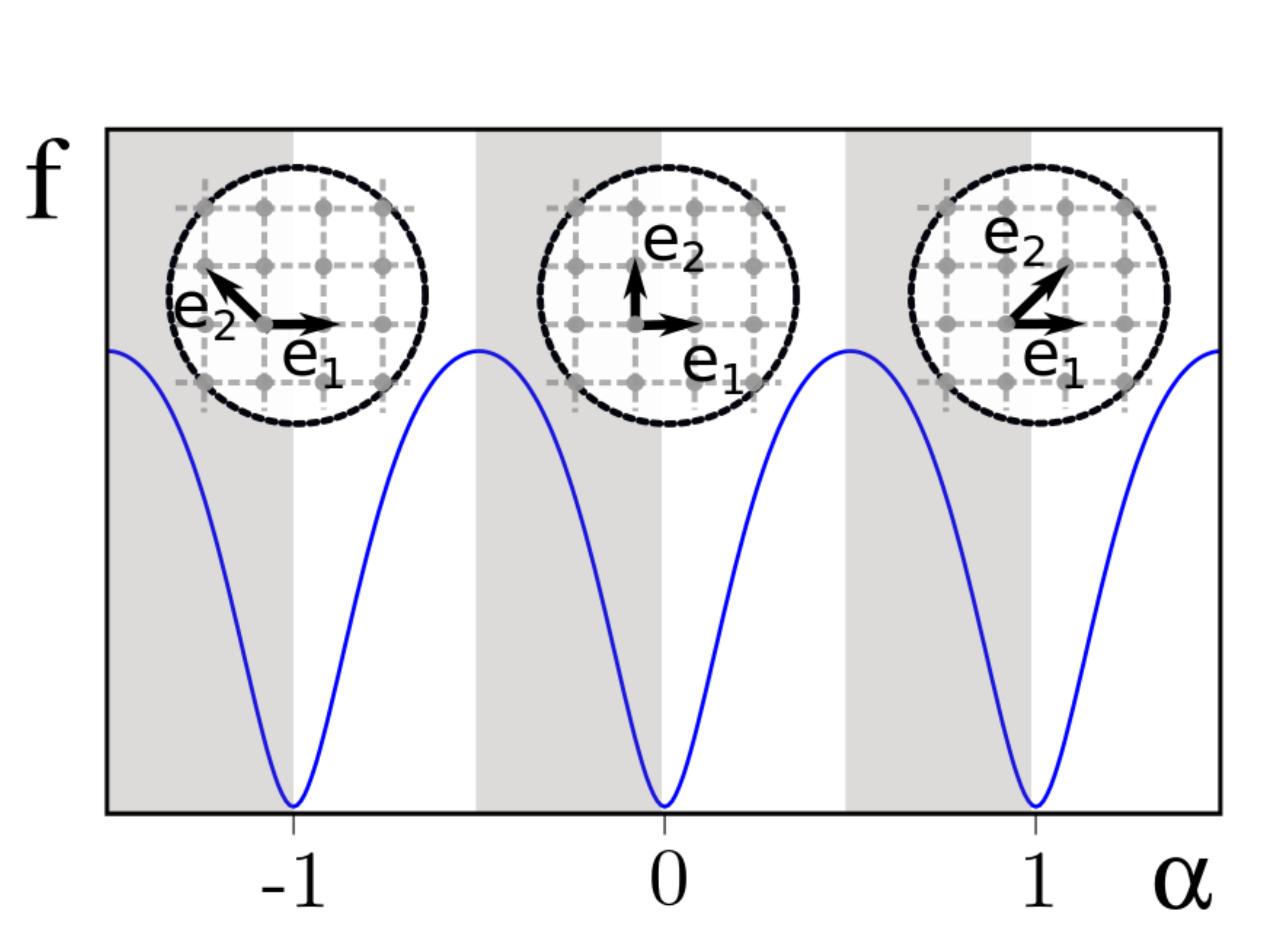}
\caption{ \small  Schematic representation of a lattice invariant shear and the associated energy barriers along the   simple shear loading path  
 $\bna\by =\mathbb{1}  +\alpha(\textbf{e}_1\otimes\textbf{e}_1^\perp)$. Alternating minimal periodicity domains are marked in gray and white.} \label{fig:shear_sch}
\end{figure}

In this Letter we propose a synthetic approach dealing with the  macroscopic quantities such as stresses and strains, while accounting correctly for  the  exact symmetry of the crystal lattice. Our main assumption is that meso-scale material elements are exposed to the  periodic   energy landscape which resolves  lattice-invariant strains, including shears related to slip \cite{bulatov1994stochastic,kaxiras1994energetics}, see Fig. \ref{fig:shear_sch}.  Our approach  follows the original proposal by Ericksen  that the energy periodicity in the space of tensors should be made compatible with geometrically nonlinear kinematics of crystal lattices \cite{Ericksen1970-rx, Ericksen1973-yt,Ericksen1977-pj,Ericksen1980-km},  and we also build upon  subsequent  important  developments of the mathematical formalism in \cite{Parry1976-zt,Folkins1991-em,Parry1998-sv,pitteri2002continuum,Conti2004-sv}.    

 This general program can be viewed as  far reaching generalization of the Frenkel-Kontorova-Peierls-Nabarro model  accounting for   energy  periodicity  along  a  single  slip plane \cite{Frenkel1939-tk,Peierls2002-xo,Nabarro2002-js}.  Scalar models with periodic energies, dealing  with  multiple slip planes,  have been  used  before to describe  dislocation cores \cite{Kovalev1993-se,Landu1994-wt,Carpio2003-yu},  to simulate dislocation nucleation  \cite{Plans2007-cx,Bonilla2007-jk,lomdahl1986dislocation,srolovitz1986dislocation}  and to capture intermittency of  plastic flows \cite{Salman2011-ij,Salman2012-oa}. Their  tensorial versions  with \emph{linearized} kinematics were considered in \cite{Minami2007-ew,Onuki2003-ln,Carpio2005-fl,Geslin2014-ad}.

In the proposed  kinematically  \emph{nonlinear} theory  the role of the order parameter is  played by the  metric tensor (characterizing local deformation), and    the  bottoms of the energy wells correspond to lattice invariant deformations. The continuum Landau  energy density of this type must  be invariant under  the infinite symmetry group $\mathrm{GL}(3,\Z)$ and since the    ground state in this case  is  necessarily hydrostatic  \cite{Ericksen1973-yt,Fonseca1987-pd},  the  regularization is necessary. Theories incorporating  various elements of the tensorial $ \mathrm{GL}(3,\Z)$  symmetry have  already proved useful in the description of  reconstructive phase transitions   \cite{Dmitriev1988-sq,Horovitz1989-lw,Sanati2001-px,Bhattacharya2004-es, Perez-Reche2009-wu,Vattre_undated-tr} and  in this Letter we extend this   idea to  the modeling of crystal  plasticity proper, see also \cite{Biscari2015-ci}.  

From the perspective of  Landau theory with an infinite number of \emph{equivalent} energy wells,  plastically deformed solid can be viewed as  a multi-phase mixture of   \emph{equivalent}   phases.  Given  the large magnitude of the 'transformation strain',   such phases  are   localized at the scale of the regularizing length and the domain boundaries appear macroscopically  as linear defects  mimicking  dislocations.  Plastic yield  can be then  interpreted as  an escape from the    reference  energy well  and   plastic mechanisms  can be  linked to  low-barrier valleys in the  energy landscape.  Friction type dissipation emerges as a result of homogenization of an overdamped athermal dynamics in a rugged energy landscape \cite{Puglisi2005-lg,Mielke2011-ck}. It is important to notice that such  approach   incorporates both  long and short range  dislocation interactions; it also correctly describes plastic slip  even though the dislocations cores are regularized and blurred on the scale of the unit cell.
  
\begin{figure}
\centering
\includegraphics[scale=0.09]{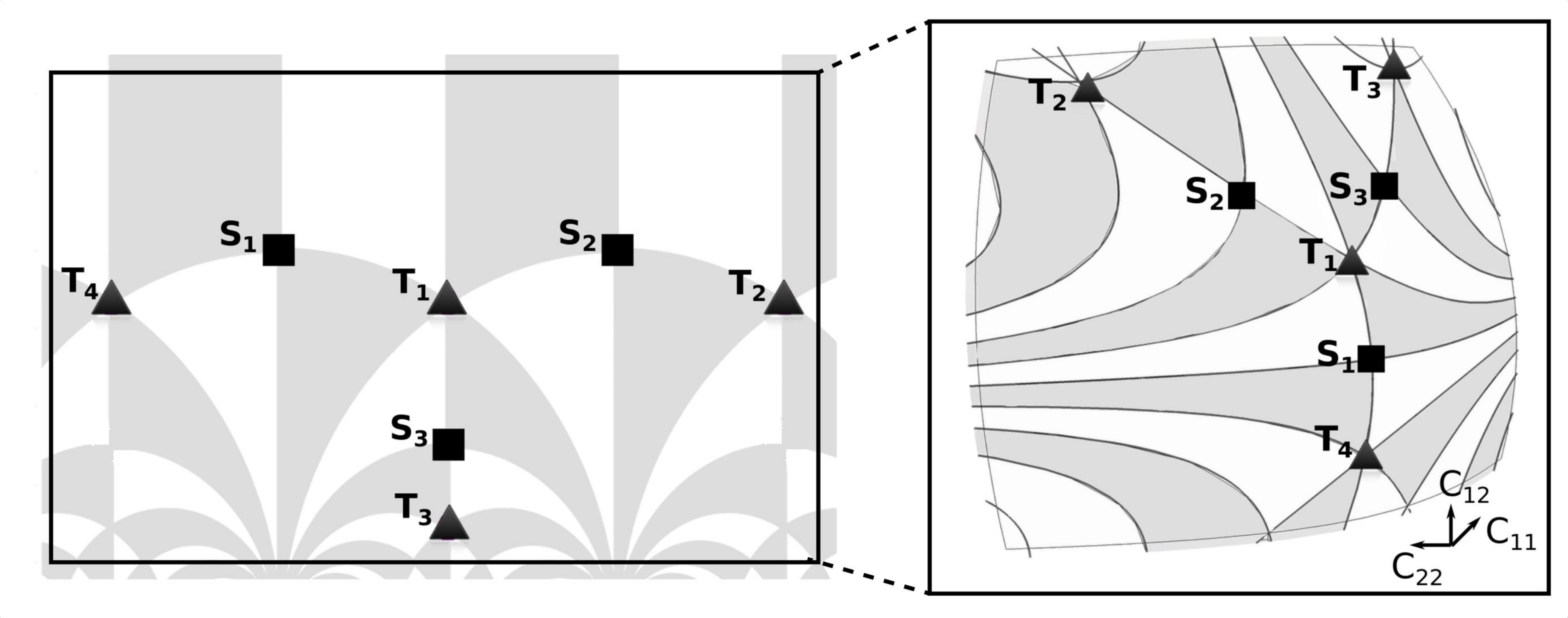}
\caption{\small  Structure of the $\mathrm{GL}(2,\Z)$ periodicity domains in the space of metric tensors: (a)  partition  of the complex  half-plane (Dedekind tessellation),    (b)  equivalent  partition  of the section $\det\bC=1$  of the space $\bC$. Points $\bf S_i$ represent the same square lattice, points $\bf T_i$ -- the same triangular lattice.}\label{fig:fig2new}
\end{figure}
  
In this Letter we use this general approach to study the peculiarities of  \emph{collective} dislocation nucleation  in crystals with different  symmetries and   show   that in low symmetry  lattices there is  a nontrivial coupling between    plastic mechanisms due to the presence of degenerate mountain passes representing  lattices with higher  symmetry. The crucial role in the formation of self induced plastic  disorder  is then  played by the unstable  high symmetry  phases, traditionally thought to be unrelated to plastic flow. The general conclusion is that,  rather  paradoxically, the  global crystal \ symmetry can  induce frustration  and become   the cause  of lattice  incompatibility  developing during plastic deformation.

Consider a  continuous deformation   $\by = \by(\bx)$, where $\by $ are actual and   $\bx$ are  reference coordinates.  The  energy density of an elastic solid   can depend on the deformation gradient $ \bna\by $  only  through  the metric tensor  $\bC=(\bna\by)^T\bna\by$.   To account for all deformations that map a  Bravais lattice into itself  we must   require that $f(\bC)=f(\bf m^T\bC \bf m)$ for  any  $\bf m$   from   a discrete group conjugate to $\mathrm{GL}$ $(3,\Z)$ and comprised of all   invertible matrices with integral entries and determinant $\pm 1$, see Supplementary Material \cite{complexn}.  In the presence of such symmetry,  the space of metric tensors $\bC$ partitions into  \emph{periodicity domains}, each one containing an  energy well equivalent to the reference one.  If  we know the structure of the energy in one of such domains,  we can use, for instance,  the  Lagrange reduction \cite{Engel1986-tm,Conti2004-sv}  to find its value in any other point. 
\begin{figure}[ht]
\centering
\includegraphics[height=2.9  cm]{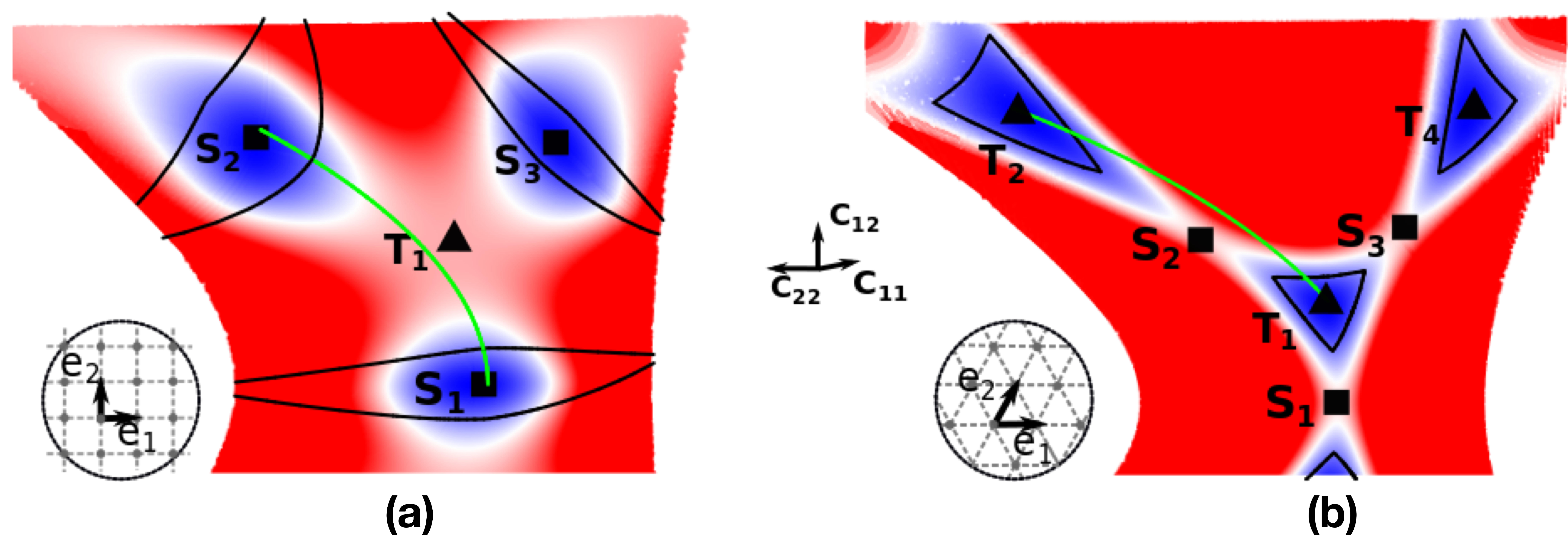} 
\caption{\small  Energy landscapes corresponding to  potential \eqref{finalenergy} with $\beta=- 1/4$ (a) and $\beta=4$ (b).  Color indicate the energy level:  blue-low, red-high.  Black lines delimit the zones of linear stability  for  the homogeneous states. Green lines correspond to the  simple shear loading paths discussed in the text.} \label{energy}
\end{figure}

In  the special case of  2D  lattices, which we focus on in what follows,  a  section $\det\bC=1$  in the  3D space of tensors $\bC$ can be  used to visualize the implied  tensorial periodicity of the energy, see Fig  \ref{fig:fig2new}.  The global picture  is made visible if we   map   this section into  a complex  half-plane  using the function  $z=C_{11}^{-1}(C_{12}+\rmi)$ \cite{Folkins1991-em,Parry1998-sv}.  For instance,   the point  $\bf S_1$ in Fig. \ref{fig:fig2new}, corresponding on the complex half plane to $ z=\rmi$  describes a \emph{square} lattice with the  basis  vectors  aligned with the close-packed directions: ${\bf e}_{1}=(1,0)$, ${\bf e}_{2}=(0,1)$. Simple shear  $ \mathbb{1}  + \,\be_1\otimes\textbf{e}_1^\perp$, where  $\textbf{e}_1^\perp$ is a vector orthogonal to  $\textbf{e}_1$,  maps this point   into its symmetric counterpart $\rmi+1$  (point $\bf S_2$);  Another square lattice, corresponding to point  $\bf S_3$ in Fig. \ref{fig:fig2new} with $ z=1/2(1+\rmi)$, can be obtained from the lattice   $\bf S_1$ by the shear $ \mathbb{1}  + \,\be_2\otimes\textbf{e}_2^\perp$.  Instead, the  point  $\bf T_1$ in  Fig  \ref{fig:fig2new}, corresponding to $z=\frac{1}{2}+\frac{\sqrt{3}}{2} \rmi $,  describes  a \emph{triangular} lattice (with hexagonal symmetry)  whose  basis vectors ${\bf e}_{1}=\gamma(1,0)$,   ${\bf e}_{2}=\gamma( 1/2, \sqrt{3}/2)$ with  $\gamma = (4/3)^{1/4}$ are  again aligned with the close-packed directions. Its closest equivalent   neighbors are   $\bf T_2$ and $\bf T_4$  corresponding to $z=\frac{1}{2}+\frac{\sqrt{3}}{2}\rmi \pm 1$. They are reachable from $\bf T_1$ by the shear deformation $ \mathbb{1}  \pm  \,\be_1\otimes\textbf{e}_1^\perp$.

To demonstrate  the   possibility  of yield-inducing instabilities in a material with such energy, it is sufficient  to  consider  a system under  the most constraining,  affine  displacement control.    Given   the gradient nature of the order parameter,  the  (spinodal)  instability of a homogeneous state  should  be linked to  the local loss of rank-one-convexity of the energy  \cite{ogden1997non,Grabovsky2014-fb}.   This is equivalent to the loss of positive definiteness of the acoustic tensor  $\bQ$ with components $  Q_{ik}=A_{ijkl}n_jn_l$, where $A_{ijkl}= \partial^2 f/(\partial_j y_{i}\partial_l y_{k})$ is the fourth-order  incremental elastic  tensor and $\bold n$ is a unit vector \cite{Van_Vliet2003-yg,Miller2008-rr}.

For illustrative purposes  we now choose a particular energy density  $f=f_v+f_d$  which   decouples   into 
a volumetric $f_v(\det\bC)$ and a deviatoric $f_d (\bC/(\det\bC)^{1/2})$  parts.  Since $\det\bC$  is invariant under $\mathrm{GL}(2,\Z)$, our symmetry constraints  concern  only the deviatoric part $f_d$. This  function needs to be specified only inside a single periodicity domain with the  suitable conditions on its boundary ensuring  required smoothness \cite{Parry1976-zt}.  The  lowest order \emph{polynomial} representation of $f_d$,  which guarantees the  continuity of the elastic moduli,  was  constructed  in \cite{Conti2004-sv}; for the  general non-polynomial representation see \cite{Folkins1991-em}.   

If  the  reference lattice  is  either square or triangular symmetry, the minimal potential can be chosen in the form \cite{Conti2004-sv}:
\begin{equation}
  f_d({\bf \tilde{C}}) =  \beta \psi_1 ( {\bf \tilde{C}}) 
 + \psi_2 ({\bf \tilde{C}})    
\label{finalenergy}
\end{equation}
where 
$
\psi_1={I_1}^4\,I_2 - 41\,{I_2}^3/99 +
7\,I_1\,I_2\,I_3/66 + {I_3}^2/1056, $ and $
\psi_2 = 4\,{I_2}^3/11  + {I_1}^3\,I_3 -  8\,I_1\,I_2\,I_3/11  +  17\,{I_3}^2/528. 
$
The hexagonal invariants here have the structure:
$
I_1 =  \frac{1}{3} (\tilde{C}_{11} + \tilde{C}_{22} - \tilde{C}_{12})$, 
$I_2= \frac{1}{4} (\tilde{C}_{11} - \tilde{C}_{22})^2 + \frac{1}{12}(\tilde{C}_{11} + \tilde{C}_{22} -
4 \tilde{C}_{12})^2$, and 
$I_3 =  (\tilde{C}_{11} - \tilde{C}_{22})^2 (\tilde{C}_{11} + \tilde{C}_{22} - 4 \tilde{C}_{12}) - \frac{1}{9} (\tilde{C}_{11} + \tilde{C}_{22} -
4 \tilde{C}_{12})^3.$  The choice  $\beta =- 1/4$   enforces the   square symmetry on the reference state, while choosing  $\beta =4$ we    bias the reference state  towards  hexagonal symmetry; the energy landscapes in  those two cases  are illustrated  in  Fig.  \ref{energy}.  The   volumetric energy density  will be chosen in the  simplest form  $f_v(s)=\mu  (s-\log(s))$,  which excludes   configurations with infinite  compression; the coefficient $\mu$ plays the role of  a bulk   modulus.
\begin{figure}[t]
\centering
\includegraphics[height=2.9 cm]{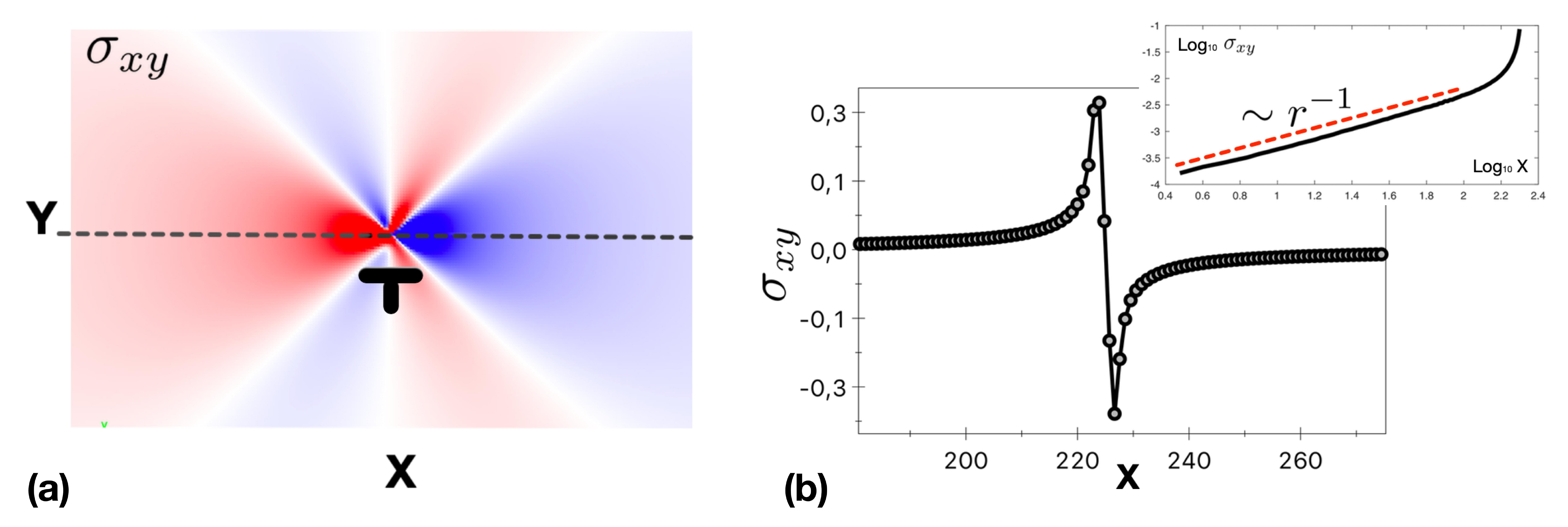}
\caption{\small  Single edge dislocation in a square lattice (interface between equivalent 'phases' $\bf S_1$ and $\bf S_2$); its Burgers vector
is horizontal,  with the  length equal to the side of the square unit cell.  (a) Finite element nodes  with color indicating the level of Cauchy stress   $\sigma_{xy}$; (b) Stress profile  along the glide plane; inset illustrates the far field asymptotics. System size $1000\times1000$. }
\label{fig:fig3}
\end{figure}

The   resulting  'yield surfaces'  are shown   in  Fig.  \ref{energy}.  To understand the nature of  the associated instabilities,  consider the case  when    a simple shear is  imposed on the boundary    
\begin{equation}
 \bna\by = \mathbb{1} +\alpha({\bf e}_{1}\otimes\textbf{e}_1^\perp).
\label{path}
\end{equation}
Starting,  in the case of square lattice,  from the  homogeneous reference state $\bf S_1$, we  find that  at instability point   the condition $\det {\bf Q}=0$  produces  two, almost simultaneously destabilized directions  $\bold q=\bna\by [\bold n]/|\bna\by [\bold n]|=(\cos\xi,\sin\xi)$: the first  one with   $\xi\approx -0.11$rad, almost perpendicular  to the deformed $\bold e_2$,   and  the second one with $\xi\approx 1.55$rad,  almost perpendicular  the the deformed $\bold e_1$.  The  near-degeneracy  of the bifurcation  is an indicator  that  two 'slip planes' may be activated.  In the case of triangular   lattice,    the  instability  along a similar loading path originating at $\bf T_1$  produces  a single  unstable direction    $\xi\approx -1.25$rad  which is  incommensurate with the lattice. In this case one can expect only one 'slip plane' to be activated. Our numerical experiments show  that the acoustic-tensor-based analytical   instability  conditions are in agreement  with direct  numerical simulations. 

Before addressing the post-bifurcational behavior  consider a single edge dislocation trapped by the lattice  somewhere far from the boundaries. In Fig. \ref{fig:fig3} we illustrate  the corresponding stress distribution which  matches the classical continuum far field with $r^{-1}$ asymptotics while also resolving (at a scale of the mesh) the core region. Solutions like this can be helpful in calibrating the   model using molecular statics simulations.

\begin{figure}
\centering
\includegraphics[height=8. cm]{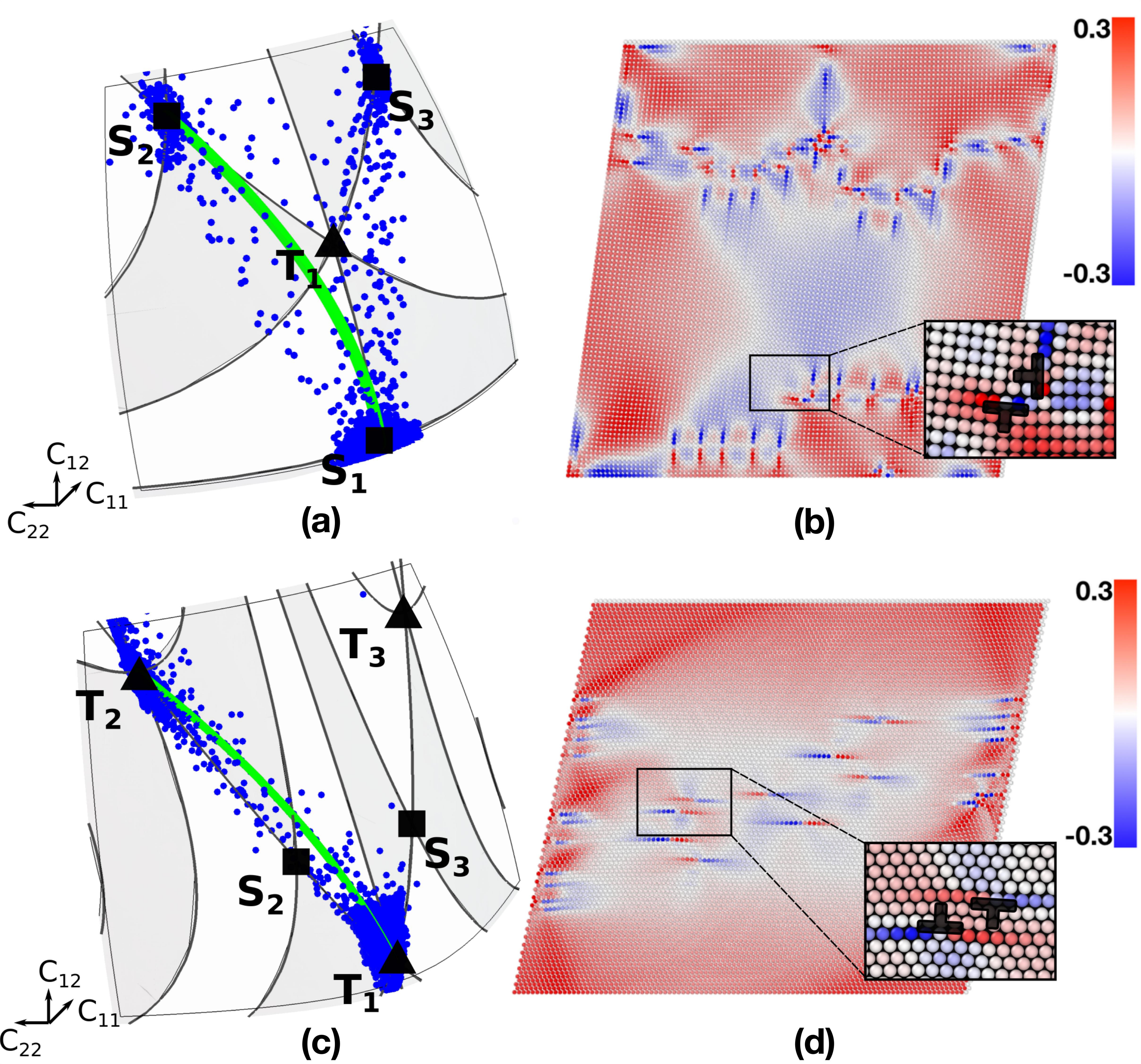}
\caption{ Collective dislocation nucleation:   (a,b)  square  lattice,  (c,d)  triangular lattice. We show two representations of the same phenomenon:   (a,c)  in the configurational space, where  green lines are simple shear  paths imposed by the loading device and blue dots indicate  the metric tensor distribution among  the elements; (b, d)  in the physical space; colors indicate  the level of  the nodal Cauchy stresses $\sigma_{xy}$. System size $200\times200$.}\label{fig:fig5new}
\end{figure}
The   collective nucleation pattern emerging after  a stress drop  is  illustrated   in Fig. \ref{fig:fig5new}  for both types of lattices.  The results are presented on both, the  configurational  space,  Fig. \ref{fig:fig5new}(a,c),  where each point corresponds to a single element of the mesh   and   the actual physical space, Figs. \ref{fig:fig5new}(b,d),  where the color of the dots (nodal points) indicate the level of stress. The configurational points,   all located initially at the bottom of the reference energy well,  disperse as a result of the massive nucleation event.   The  ensuing  spatial dislocation distribution is  quasi-regular with  pile-ups at  the  rigid boundaries. Note the  formation of  characteristic entanglements  with dislocations on two slip planes blocking each other (in the case of square lattice);   there is also some disorder due to  unavoidable numerical noise.

Note that in the case of square lattice,  the system  is driven  by the loading device from the  reference state $\bf S_1$ towards the equivalent state $\bf S_2$.  At the "yielding"  threshold,   which marks  the end of the elastic regime,  the homogeneous configuration $\bf S_1$  loses stability  and  the ensuing  pattern represents   (outside  the core regions) a mixture of three  'pure' states   $\bf S_1$, $\bf S_2$ and $\bf S_3$, see Fig.\ref{fig:fig5new}(a).  While the appearance of the state $\bf S_2$ is natural, because the corresponding  'plastic mechanism' is   favored   by the loading, the main complexity of the resulting dislocation pattern is due to the emergence of the state $\bf S_3$.  It indicates the activation of the second   plastic mechanism, decoupled (in the nonlinear theory) from the  first one. 

The appearance of the state  $\bf S_3$ can be understood if we recall that the linear stability analysis  predicted two almost simultaneously unstable modes aligned with the slip  directions in the deformed state.  While one of these directions is indeed aiming towards the energy well $\bf S_2$, the other one, which  bifurcates first,  is   directed   towards $\bf S_3$.  Our  numerical simulations show that the latter instability  mode grows faster  which can be interpreted as, somewhat counter-intuitive, early stage  dominance  of  the secondary  'plastic mechanism'. The flow of  configurational points   passes near the  unstable equilibrium state  $\bf T_1$, corresponding to a triangular lattice, where it splits into three   streams directed towards the  configurations $\bf S_1$, $\bf S_2$ and $\bf S_3$, see Movie S1 in \cite{complexn}.  

\begin{figure}
\centering
\includegraphics[scale=0.2]{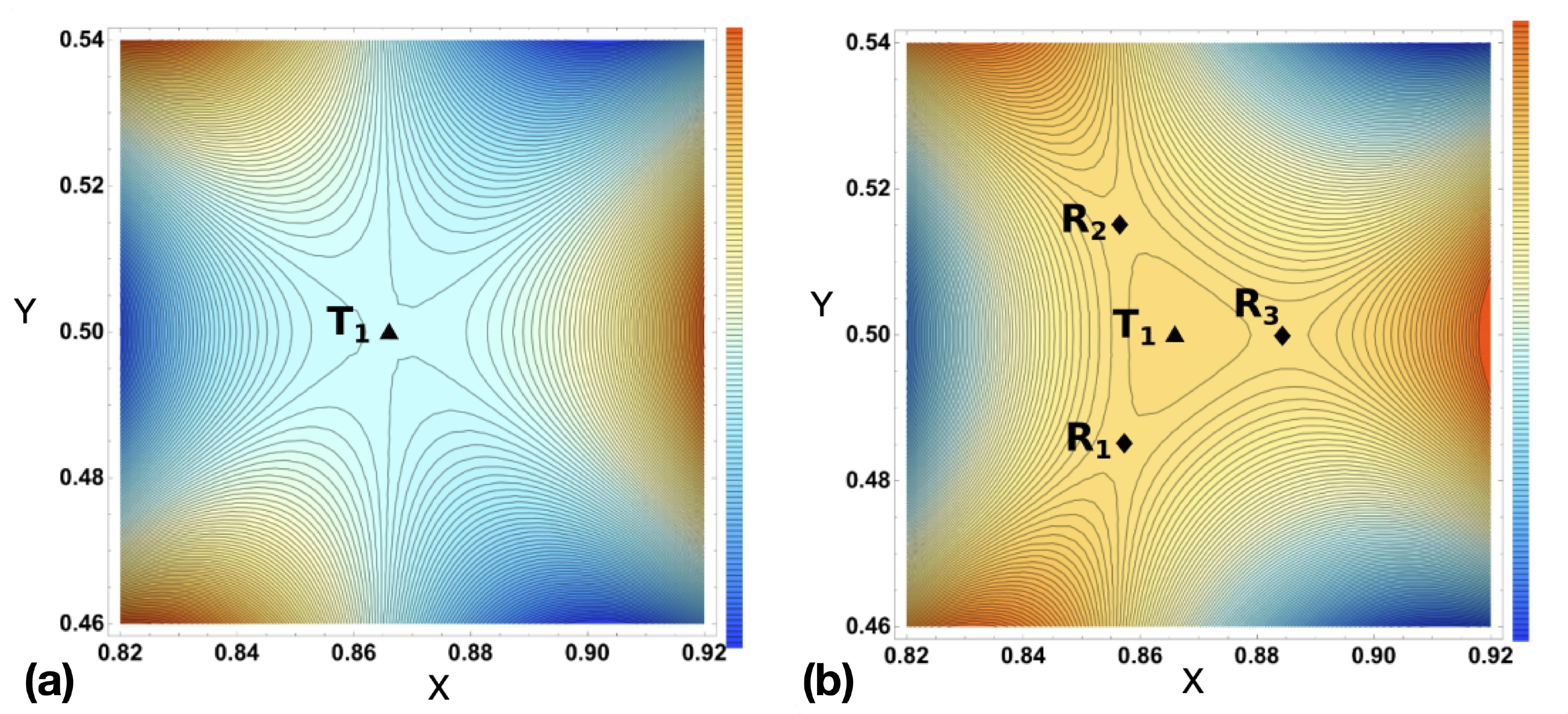}
\caption{Level sets of the  energy density on the surface $\det\bold C=1$ around the point $\bf T_1$: we use the parametrization  $ C_{11}= 1/Y$, $  C_{22}= X^2+Y^2/Y$,  $C_{12}=X/Y$;   (a) Klein invariant based  potential;  (b) polynomial potential \eqref{finalenergy} with $\beta =- 1/4. $}\label{fig:fig6}
\end{figure}

This behavior  becomes more transparent if we consider a  smoother  energy potential.  Note  that the  symmetry transformations from $\mathrm{GL}(2,\Z)$ correspond on  the upper complex half-plane to  the   fractional (Moebius) transformations  with integral entries of the type $  \big(m_{22}z+m_{12}\big)\big/\big(m_{21}z+m_{11}\big)$   \cite{Parry1976-zt,Folkins1991-em}.   This observation links  the infinitely periodic energy densities for 2D crystalline materials with  the classical modular functions \cite{schoeneberg2012elliptic}, with the most well known example provided by  the Klein invariant $J(z)$ \cite{van1953coefficients}, see  Supplementary Material for more details \cite{complexn}. One can show that for this   holomorphic function  $J|_{\bf S_i}=1$, $J'|_{\bf S_i}=0$,  while  $J|_{\bf T_i}=J'|_{\bf T_i}=J''|_{\bf T_i}=0$.  Therefore, the   corresponding potentials   with the reference   square and   triangular lattices  can be chosen in the form:  $f_d(z)=|J(z)-1|$ (square lattice) and $f_d(z)=|J(z)|^{2/3}$ (triangular lattice); the exponents are  chosen to ensure   a non-degenerate  linear-elastic response  close to the bottoms of the energy wells.  The  energy landscapes and the yield surfaces for such  potentials  are qualitatively similar to the ones presented  in Fig. \ref{energy}.

Note that the choice $f_d(z)=|J(z)-1|$  for a square lattice  turns  the 'triangular'  critical point  $\bf T_1$ and all its symmetric counterparts into degenerate `monkey saddles',     characterized by  the local Taylor expansion  of the form  $ x^3 - 3 xy^2$, see Fig.\ref{fig:fig6}(a).  The flow of configurational points directed initially towards such unstable state (say, $\bf T_1$)  will therefore  necessarily split into three streams directed towards the stable states (say, $\bf S_1$, $\bf S_2$ and $\bf S_3$).  

 Superficially, the situation looks   a bit different  in the case of the polynomial energy  \eqref{finalenergy},    where the Hessian is nondegenerate at the point  $\bf T_1$ which corresponds in this case to a shallow  energy maximum.  However this maximum is surrounded by the three nondegenerate saddles $\bf R_1$, $\bf R_2$ and $\bf R_3$ describing  rhombic  lattices, see Fig.\ref{fig:fig6}(b), and the general conclusion about the activation of the secondary plastic mechanism and the ultimate dispersion over three energy wells  $\bf S_1$, $\bf S_2$ and $\bf S_3$ remains valid. Note that the implied  coupling of the plastic mechanisms  would have to be \emph{postulated} in the phenomenological plasticity theory. 
 
 The picture is simpler  in the case of  a triangular lattice  where  the loading  \eqref{path}  from $ \bf T_1$ to  $ \bf T_2$ produces   a mixture of only two  'pure' states   $ \bf T_1$, $ \bf T_2$, see Fig.\ref{fig:fig5new}(d). The latter  can be interpreted as the activation of a single plastic mechanism, the one  favored by the loading, see Movie S2 in \cite{complexn}.

To conclude, our model shows that crystal plasticity naturally arises from
nonlinear elasticity, if the   tensorial symmetry of the crystal lattice is properly
accounted for.  The  memory of the  atomic lattice  in such  infinitely periodic   Landau  theory  is present in the form of  the information about the  affine mappings  that leave the energy density  invariant.  Athermal evolution in the  regularized  model  of this type  can lead to  temporal and spatial complexity which our analysis predicts  to be highly sensitive to both, the crystallographic symmetry and the orientation of the crystal  \cite{weiss2015mild,Sparks2018-zp}. In particular, our study highlights the  crucial   role played in  plastic deformation by the  degenerate saddle points  of the  energy  representing seemingly irrelevant, unstable crystallographic  phases;  for similar effects  in   other fields see \cite{rehbein2011we,wales2003energy,shtyk2017electrons,bociort2005generating}. More generally, the proposed  Landau theory perspective  on crystal plasticity promises to become an important new tool in the study of inelasticity  at the micro/nano scales  where the conventional  engineering theories fail to access  strength, account for fluctuations and adequately describe size effects \cite{greer2011plasticity,zhang2016approaching,zhang2017taming}.

\end{document}